\documentclass[%
preprint,
superscriptaddress,
showpacs,
showkeys,
amsmath,amssymb,
aps,
prb,
]{revtex4-2}
\usepackage{graphicx,amssymb,amsmath,epsfig}
\usepackage{color}
\usepackage{comment}
\usepackage[usenames,dvipsnames,svgnames,table]{xcolor}
\usepackage{babel}
\usepackage[lofdepth,lotdepth,caption=false]{subfig}
\usepackage[pdfstartview=FitH]{hyperref}
\usepackage{hyperref}
\newsavebox{\bigleftbox}
\hypersetup{
 colorlinks=true,
 citecolor=Blue,
 linkcolor=Blue,
 urlcolor=Blue
}
\usepackage[normalem]{ulem}
\usepackage{breqn}

\begin{document}

\title{%
Length and torsion dependence of thermal conductivity in twisted graphene nanoribbons
} 

\author{Alexandre F. Fonseca}
\email[E-mail me at: ]{afonseca@ifi.unicamp.br}
\affiliation{Universidade Estadual de Campinas$,$ Instituto de F\'{i}sica Gleb Wataghin$,$ Departamento de F\'{i}sica Aplicada$,$ 13083-859$,$ Campinas$,$ SP$,$ Brazil}
\affiliation{Division of Engineering and Applied Science$,$ California Institute of Technology$,$ Pasadena$,$ CA$,$ 91125$,$ USA}

\author{Luiz Felipe C. Pereira}
\affiliation{Departamento de F\'{\i}sica$,$ Universidade Federal de Pernambuco$,$ 50670-901$,$ Recife$,$ PE$,$ Brazil}
\affiliation{Dipartimento di Fisica$,$ Sapienza Universit\`a di Roma$,$ Roma$,$ 00185$,$ Italy}

\date{\today}

\begin{abstract}
Research on the physical properties of materials at the nanoscale is
crucial for the development of breakthrough nanotechnologies.  One of
the key properties to consider is the ability to conduct heat, i.e.,
its thermal conductivity.  Graphene is a remarkable nanostructure with
exceptional physical properties, including one of the highest thermal
conductivities (TC) ever measured. Graphene nanoribbons (GNRs) share
most fundamental properties with graphene, with the added benefit of
having a controllable electronic bandgap. One method to achieve such
control is by twisting the GNR, which can tailor its electronic
properties, as well as change their TC. Here, we revisit the
dependence of the TC of twisted GNRs (TGNRs) on the number of applied
turns to the GNR by calculating more precise and mathematically well
defined geometric parameters related to the TGNR shape, namely, its
\emph{twist} and \emph{writhe}. We show that the dependence of the TC
on \emph{twist} is not a simple function of the number of turns
initially applied to a straight GNR. In fact, we show that the TC of
TGNRs requires at least two parameters to be properly described.  Our
conclusions are supported by atomistic molecular dynamics simulations
to obtain the TC of suspended TGNRs prepared under different values of
initially applied turns and different sizes of their suspended part.
Among possible choices of parameter pairs, we show that TC can be
appropriately described by the initial number of turns and the initial
twist density of the TGNRs.
\end{abstract}



\maketitle

\section{Introduction}
\label{intro}
Twist in one-dimensional materials can be either a hindrance or an
advantage.  It could be a problem when dealing, for example, with the
installation of long cables~\cite{sun98}, disentangling twisted
headphone wires or simply washing the garden or the car with a
hose~\cite{goriely97}.  However, it could be useful when extracting
elastic parameters of a nanowire~\cite{fonseca2006}, setting up
helical artificial muscles~\cite{ray2012,ray2014} or developing
torsional-based elastocaloric refrigerators~\cite{ray2019}.  Knowing
the relation between twist and physical properties of filaments in
general is important for solving problems in several areas ranging
from engineering~\cite{sun98} to biomedicine~\cite{hornig2010} and
molecular
biology~\cite{bauerDNA1993,kleninDNA2000,barbiChromatin2012}.

In the particular case of graphene nanoribbons (GNRs), the effects of
twisting on their properties have been predicted to be useful in
applications such as sensors and
switches~\cite{yakobsAPL2011,liPE2012,AlCarbon2013}.  As the term
``twisted graphene'' became usual to describe the relative rotation of
one graphene layer with respect to the other in bilayer graphene
structures, it is important to make clear that in this work, the words
``twist'' or ``torsion'', as well as the term ``twisted GNRs''
(hereafter referred as TGNRs for short) means the application of twist
or torsion along the longitudinal axis of a single GNR.

Gunlycke {\it et al.}~\cite{gunNanoLett2010} showed that edge
termination can induce twisting in GNRs (at least in the case of small
width GNRs) and that TGNRs present different band-gap behavior when
compared to flat and straight GNRs.  Sadrzadeh, Hua and
Yakobson~\cite{yakobsAPL2011} showed that hydrogen-terminated
armchair-edge GNRs present a twist dependent band-gap.
Koskinen~\cite{koshAPL2011} demonstrated a certain equivalence between
the effects of tensile and twisting strains on the electronic
structure of GNRs.  Tang {\it et al.}~\cite{tangAPL2012}, Li {\it et
  al.}~\cite{liPE2012} and Xu {\it et al.}~\cite{xuSSC2015}
investigated metallic-to-semiconductor transitions in armchair- and
zigzag-edge TGNRs, while several other studies also confirmed the
dependence of electronic and magnetic properties of GNRs on
longitudinal twist, and even suggested applications
~\cite{yueCarbon2014,jiaCarbon2014,BoZhangPRL214,bretimCarbon2019,thakurPE2020}.

Mechanical properties of TGNRs have also been studied.
Li~\cite{liJPhysD2010} investigated the stretchability of TGNRs.
Cranford and Buehler~\cite{cranbuehMS2011} presented a comprehensive
mechanical study of TGNRs including their conversion to helical GNRs.
Dontsova and Dumitric{\u a}~\cite{dumiJPCL2013} investigated the
mechanics of twisted single and few-layer GNRs.
Diniz~\cite{dinizAPL2014} and Xia {\it et al.}~\cite{xiaPCCP2016}
studied the structural stability of TGNRs while Savin, Korznikova and
Dmitriev~\cite{saviMechMat2019} showed that TGNRs have larger bending
stiffness than flat ones.  Further studies demonstrated that the
application of large amounts of twist can lead to the formation of GNR
scrolls and supercoils ~\cite{shahabiSciRep2014,savinEPL2020}, the
formation of helical ribbons~\cite{nikiJPCL2014}, changes in the
strength of TGNRs with grain boundaries ~\cite{liuPCCP2015}, and the
localization of twisting as topological solitons on substrates
~\cite{savinPRB2020}.

The lattice thermal conductivity (or simply ``TC'' from now on) of
graphene has been extensively studied so far (see, for example,
Refs.~\cite{Luiz2013,Xu2014,FanNL2017,FanPRB2017}).  Nonetheless, very
few studies have addressed the dependence of the TC of GNRs on the
amount of twist
~\cite{shenMolPHys2014,weiJAP2014,ruraliJPD2017,terada2020,savinEJPA2020,liu2021PCCP}. Most
of these studies show that increasing the amount of twist decreases
the TC, although one of them~\cite{savinEJPA2020} found an inverse
behavior arguing that twist increases the local tension strain, and
thus the contribution of the acoustic out-of plane phonon modes to the
TC of the TGNR.

There are even fewer experiments with TGNRs. Chamberlain {\it et
  al.}~\cite{chambeACS2012} used carbon nanotubes as nanoreactors to
assembly and produce sulfur-terminated GNRs, including TGNRs.  Cao
{\it et al.}~\cite{caoAPL2013} obtained TGNRs or curled GNRs by
thermal annealing poly-methyl methacrylate (PMMA) terminated GNRs, and
Jarrahi {\it et al.}~\cite{zarrahiNANOSCALE2013} studied their
photoresponse.  An important observation is that transmission and
scanning electronic micrographs of TGNRs in those references reveal
that TGNR structures are not regularly twisted GNRs as those
considered in most of the previously cited modeling and simulation
works.  In these studies, only one parameter was considered to
characterize the TGNR geometry: the initial number of turns applied to
its axis.  However, the TGNR properties might be dependent also on the
GNR length and the curls and folds that form due to its low flexural
rigidity and thermal fluctuations, as seen in the experimental
micrographs.  A study of the TC of TGNRs that takes into account these
features is missing and can reveal a higher level of complexity, which
would be required for further development of precise applications.

Two important questions arise from the above discussion: (i) how to
precisely define and determine the geometric features of a TGNR at
finite temperature?  and (ii) how to describe the dependence of the TC
of a TGNR on these geometric features, including twist and length, at
finite temperatures?  In the present study, we are going to answer
both questions.

Recently, one of us~\cite{fonsecaPRB2021} developed a method to
precisely calculate the geometric features of a TGNR suspended by two
substrates.  It was demonstrated that the degree of twist (in a more
precise mathematical sense) of a given TGNR is not solely dependent on
the number of turns initially applied to it, but also on the size of
its suspended portion.  One reason for this is that the TGNR's
extremities lay on the substrates, becoming flat and not contributing
to its total twist.  As a result, the initial turns applied to the
TGNR axis become more densely distributed along the suspended part.
This increases the twist density of the TGNR favoring the so-called
{\it twist-to-writhe} conversion (TWC)~\cite{fuller1971,goriely2000}
phenomenon (see the detailed description ahead in section \ref{A})
which allows part of the torsional stress in a ribbon to be released
by flexural deformations of the TGNR axis.

Furthermore, Fonseca~\cite{fonsecaPRB2021} used these features to
propose that the total twist of a TGNR can be tuned by simply changing
the distance between the substrates holding its ends.  He showed that
the total (real) twist of a TGNR can be changed without adding or
removing torsion/rotation at the ends of the TGNR.  One advantage of
this method is to provide a more precise way to determine the total
twist of a TGNR and, then, correlate it to other physical properties.
Since the literature is mostly limited to the prediction of physical
properties of regularly twisted GNRs, here we explore the above
geometric features of suspended TGNRs, and their dependence on the
size of their suspended part, to investigate the dependence of the
TGNR's TC on the total amount of twist, the size of the suspended part
and other TGNR's geometric parameters.  We show that the TC of TGNRs
cannot be fully determined by a unique geometric parameter, and that
it requires, at least, two parameters.

In the next sections, we present the theoretical background for
calculating the total geometric twist of a piece of TGNR, and the
computational methods employed for the calculation of the TC.  Then,
we present our results and discussions, followed by our conclusions.

\section{Theory and Methodology}

\subsection{Geometric parameters of a TGNR} \label{A}
The {\it twist-to-writhe} conversion (TWC) phenomenon, mentioned in
the \hyperref[intro]{Introduction}, is well known in twisted
filamentary structures~\cite{fuller1971,goriely2000}. It consists of
releasing a rod's torsional stress by spontaneous bending and folding,
after the twist density reaches a critical value. This TWC has been
shown to satisfy the C{\u a}lug{\u a}reanu-White-Fuller {\it linking
  number} ($Lk$) theorem~\cite{goriely2000,caluga61,fuller78}:
\begin{equation}
    \label{lk}
    Lk = Tw + Wr \, 
\end{equation}
where $Tw$ and $Wr$ are the total (real) amount of twist and a
quantity called {\it writhe} of a curve which measures its
non-planarity, respectively.  The linking number, $Lk$, is a geometric
parameter of a pair of closed curves and although it is well defined
in terms of a double integral along them, it has been shown to be an
integer equal to half the number of times one curve crosses the
other~\cite{berger2006}.

The total twist, $Tw$, of a pair of curves and the writhe, $Wr$, of
one space curve, are given by the following integrals along the
corresponding curves~\cite{berger2006}:
\begin{equation}
    \label{tw} 
    Tw=\frac{1}{2\pi}\int_{\textbf{x}} \textbf{t}_{\textbf{x}} \cdot 
    \left( \textbf{u} \times \frac{d\textbf{u}}{ds} \right)ds \, ,
\end{equation}
\begin{equation}
    \label{wr}
    Wr=\frac{1}{4\pi}\int_{\textbf{x}} \int_{\textbf{x}} 
    \frac{(\textbf{t}_{\textbf{x}(s)} \times \textbf{t}_{\textbf{x}(s')})\cdot
    (\textbf{x}(s)-\textbf{x}(s'))}{|\textbf{x}(s)-\textbf{x}(s')|}dsds' \, ,
\end{equation}
where $s$ is the arc-length of the curve $\textbf{x}$, $\textbf{t}$ is
its unitary tangent vector and $\textbf{u}$ is a unitary vector
orthogonal to $\textbf{t}$ and pointing from the curve $\textbf{x}$ to
its parallel curve.  All these vector quantities are functions of
$s$. The total length of the curve $\textbf{x}$ is simply given by
$L=\int_0^L ds$.

Suppose we initially prepare two space closed curves such as to
present a certain amount of $Lk$. The C{\u a}lug{\u
  a}reanu-White-Fuller theorem guarantees that $Lk$ is always
conserved no matter how the curves change along the time, provided
they remain closed.  Changes to the curves mean changes to their
values of $Tw$ and $Wr$ through equations (\ref{tw}) and (\ref{wr}),
respectively. According to the theorem, these changes are such that
$Tw+Wr$ remains constant as along as the curves remain closed. An
interesting feature is that the theorem has also been shown to hold
for a pair of non-closed curves if their extremities are flat and
belong to the same plane~\cite{berger2006}.  There exists a pair of
parallel open curves with $Tw=Wr=0$ that connect the first two ones at
infinity~\cite{berger2006}.  A similar argument can be made for a pair
of open curves having a semi-integer value of $Lk$.  As long as the
ends of the curves lay on parallel planes, there exists a pair of
twisted parallel curves with $Wr=0$ and $Tw=0.5$ that connect the
first pair at infinity.

The two space curves required to calculate the twist and writhe of our
TGRNs can be defined prior to applying the initial turns to the
straight GNR. The first curve can be the GNR axis, and the second
curve can be a line parallel to the first.  Numerically, both curves
can be defined by sets of co-linear carbon atom positions along the
main length of the straight untwisted GNR.  Once the curves are
defined, if one fixes one end of a GNR and applies $n$ turns to the
other end while keeping the GNR straight, the twist will be $Tw=n$ and
the writhe $Wr=0$.  Thus, the initial linking number applied to the,
now twisted, GNR is $Lk = n$.  If the ends of this TGNR are placed
such that they belong to the same plane and are not allowed to rotate
back to release the initially applied torsional stress, the value of
$Lk$ will remain constant.

As shown by Fonseca~\cite{fonsecaPRB2021}, if the extremities of a
TGNR are laid down on two different planar substrates, and their
planes coincide, the $Lk$ theorem can be applied to the TGNR and
eqs. (\ref{tw}) and (\ref{wr}) can be used to infer the values of $Tw$
and $Wr$ of the suspended part.  Additionally, it is possible to
investigate how $Tw$ and $Wr$ vary with several other parameters and
physical conditions, such as the distance between the substrates,
temperature, etc.  The interesting thing is that as long as the
extremities of the TGNR are kept on the substrates (and van der Waals
forces guarantee that), no matter how $Tw$ and $Wr$ change with other
physical conditions, $Lk$ will remain the same.  For instance,
Fonseca~\cite{fonsecaPRB2021} showed that it holds true for changing
the distance between substrates and the temperature of the system.

Here, we will investigate how the TC of TGNRs depends on $Lk$, its
length, $d$, as well as its $Tw$ and $Wr$ taking into account that
these last two quantities change with both $Lk$ and $d$.  We will
analyze the twist and writhe of non-closed TGNRs to which an integer
or semi-integer number of turns was initially applied.

\subsection{Computational methods}

The TC calculation for the TGNRs will be performed by non-equilibrium
molecular dynamics (NEMD) simulations using the Adaptive
Intermolecular Reactive Empirical Bond Order (AIREBO)
potential~\cite{brenner2022,stuart200} as implemented in
LAMMPS~\cite{Plimpton2022}.  AIREBO is an extension of the REBO
potential originally developed by Brenner \textit{et
  al.}~\cite{brenner2022}, which includes Lennard Jones and torsional
potential terms~\cite{stuart200}.  After more than two decades, the
AIREBO potential is still being successfully used to simulate
structural~\cite{ruoffSCIENCE2010,LeviJCCP2020,guilhermeCTrends2022,guilhermeMRSA2023}
and thermal
properties~\cite{fonsecaMuniz2015,lisenkov2016elastocaloric,CantuarioANNPHY2019,LiSCICHINE2020,zhao2022room,TatianaPRB2022,ribeiro2023elastocaloric}
of carbon nanostructures, including heat transport
simulations~\cite{Luiz2013,Xu2014,FanNL2017,FanPRB2017,he2015JNN,ito2022AIPAdv}.
One fundamental aspect of our choice is the computational time
involved.

Nonetheless, we must keep in mind that AIREBO does not quantitatively
reproduce absolute TC values for carbon nanostructures. In order to
overcome this limitation, we will focus on how the TC of TGNRs depends
on their geometric features, and not on its absolute value. Zhang and
collaborators, for example, have performed a similar study using the
original REBO/AIREBO to investigate TC trends in graphene with the
number of isotopes~\cite{fonsecaKJ2010}, and in graphene oxide with
the percentage of oxygen coverage~\cite{fonsecaKJ2014}.

The TC simulation protocol can be described as follows.  TGNRs having
$Lk=0$ (non-twisted and straight), $0.5, 1.0, 1.5$ and $2.0$ are
generated by fixing one of their ends and applying $2\pi Lk$ rads with
respect to the ribbon axis to the opposite end.  With the extremities
fixed, the TGNRs are optimized with the conjugate gradient energy
minimization algorithm as implemented in LAMMPS (with energy and force
tolerances of $10^{-8}$ eV and $10^{-8}$ eV/\AA, respectively).  Then,
the TGNR extremities are placed at $\sim 3.4$ \AA\mbox{} of distance
to two different substrates modeled as large area square shape
graphene single layers of $\sim287$ \AA\mbox{} of side, distanced by
$d$.  The amount of area of the TGNR extremities laid on each
substrate is such that its suspended part has one of the following
sizes: $d=100, 200, 300, 400$ and $500$ \AA. Each TGNR is further
equilibrated at 300 K for about 1 ns using a Langevin thermostat, with
0.5 fs as time step and 1 ps as thermostat damping factor.  Long time
simulations are required in order to guarantee that the suspended part
of the TGNR reaches equilibrium.  During these simulations, the
substrates are kept fixed and the objective of this part of the
protocol is to get the equilibrium shape of the TGNR at the chosen 300
K temperature. From the equilibrium shapes of TGNRs, $Tw$ and $Wr$ can
be calculated using the algorithm described in
Ref.~\cite{fonsecaPRB2021}.

Armchair GNRs of about 600 \AA\mbox{} length and 33 \AA\mbox{} width
are considered in the present study.  They are fully
hydrogen-passivated.  As classical MD simulations show no special
dependence of TC with the direction of thermal conduction in pristine
graphene we have not repeated the simulations with zigzag GNRs
\cite{Luiz2013}.

The TC was calculated as follows.  In a real situation, the substrates
play the role of thermal baths. However, the simulations to determine
the TC of TGNRs will be performed in the nanoribbons alone, without
the substrates, to save time in the thermal equilibration of the
substrates and to avoid the unknown heat transmission and thermal
resistance at the interface between the substrate and the TGNR.
Thermostats at $T_{\mbox{\scriptsize{HOT}}}=350$ K and
$T_{\mbox{\scriptsize{COLD}}}=250$ K are, then, applied to the carbon
atoms that, in the simulations with substrates, laid on each of them.
In the absence of the substrates, a free TGNR would rotate and release
its torsional stress. In order to avoid that, the hydrogen atoms that
also laid on the substrates are kept fixed during the simulations to
determine the TC. The carbon and hydrogen atoms that are suspended in
the simulations with the substrates, are allowed to freely evolve,
i. e., no thermostat or constraints are applied to them.  The
simulations to determine the TC of each TGNR were performed for, at
least, 40 ns.

\begin{figure}[htb]
\centering
\includegraphics[width=0.9\linewidth]{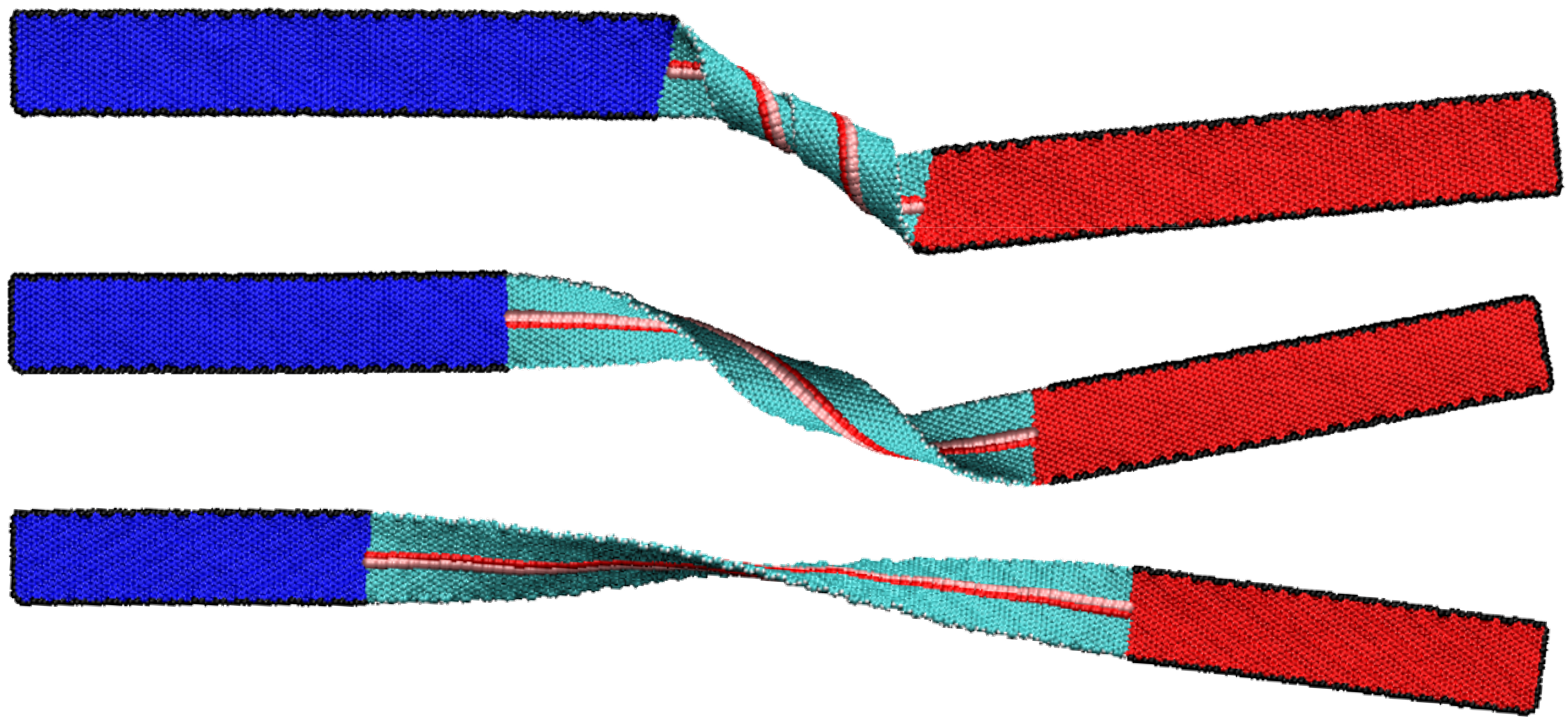}
\caption{Examples of atomistic structures of TGNRs with $Lk=2$ and
  $d=100$ \AA\mbox{} (top), $Lk=1$ and $d=200$ \AA\mbox{} (middle),
  and $Lk=0.5$ and $d=300$ \AA\mbox{} (bottom). Red (blue) carbon
  atoms represent those to which thermostats at
  $T_{\mbox{\scriptsize{HOT}}}=350$ K
  ($T_{\mbox{\scriptsize{COLD}}}=250$ K) were attached in the
  simulations to determine the TGNRs' TC.  Black, white and cyan atoms
  correspond to fixed hydrogen, free hydrogen and free carbon atoms,
  respectively, during those simulations. Red and pink lines of atoms
  define the two space curves representing the geometry of the
  suspended part of the TGNR.}
\label{fig01}
\end{figure}

Figure \ref{fig01} depicts three TGNRs with different values of $Lk$
and $d$. There, red and blue atoms are thermostated at
$T_{\mbox{\scriptsize{HOT}}}$ and $T_{\mbox{\scriptsize{COLD}}}$
temperatures, respectively, while black atoms are kept fixed. Cyan,
white, red and pink atoms at the suspended part of the TGNR are
allowed to evolve freely. Red and pink atoms in the suspended part of
the TGNR are those whose coordinates will be used to obtain the space
curves needed to calculate $Tw$ and $Wr$ using equations (\ref{tw})
(\ref{wr}), respectively.  For every TGNR, the TC, $Tw$ and $Wr$ of
the suspended part were calculated.

\subsection{Theoretical method to determine the TC}

\begin{figure}[htb]
\centering
\includegraphics[width=0.9\linewidth]{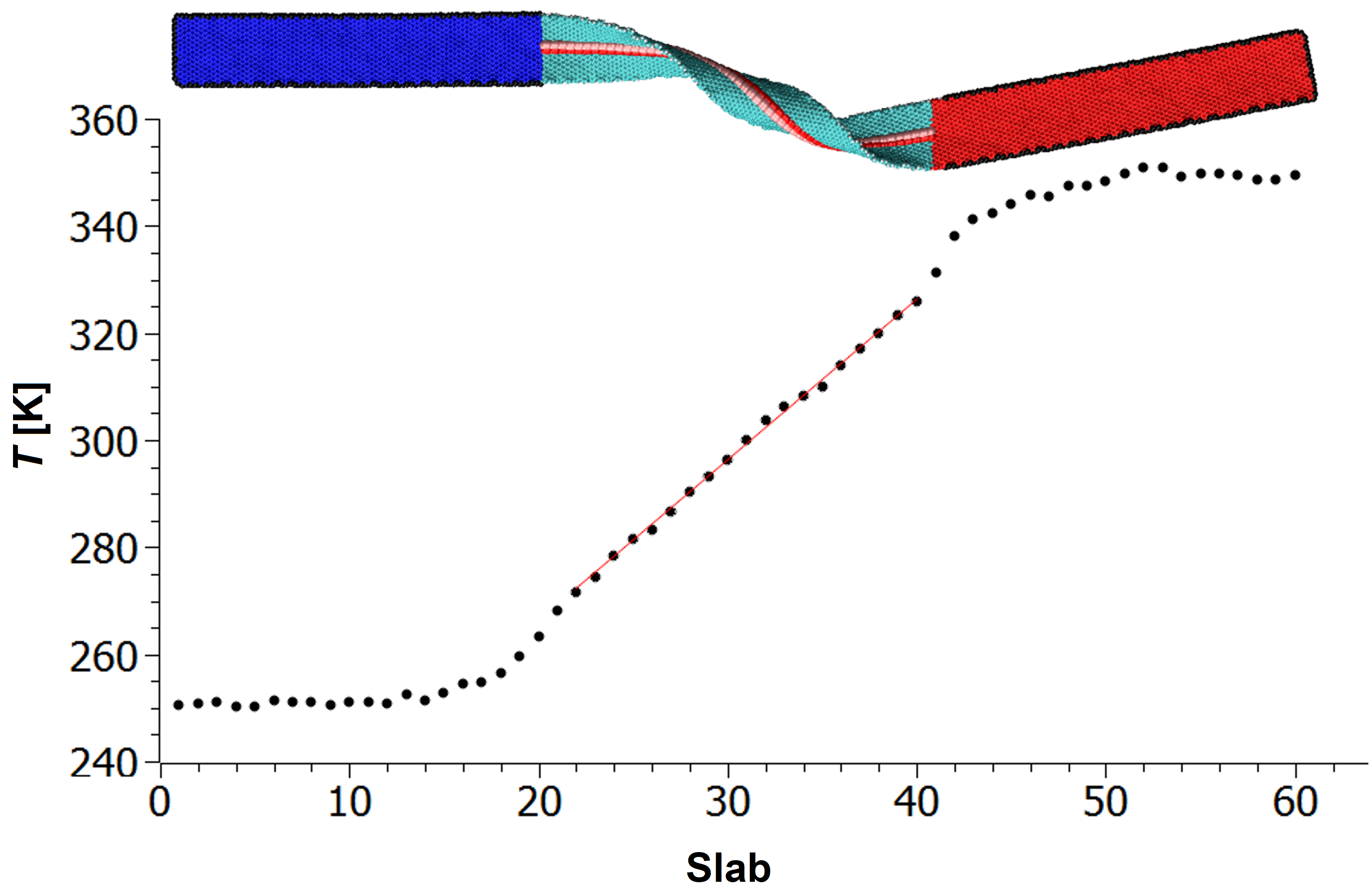}
\caption{A typical example of the steady state temperature profile
  along the TGNR of the middle panel of Fig. \ref{fig01}, after 40 ns
  of a NEMD simulation to determine its TC. Each point corresponds to
  a slab along the TGNR. The line connecting the points at the central
  region is a fitting straight curve needed to obtain the temperature
  gradient. The meaning of the colors of the atoms is the same as
  given in the caption of Fig. \ref{fig01}.}
\label{fig02}
\end{figure}

The TC, $\kappa$, of a system along a direction $\mathbf{x}$, can be
obtained from Fourier law:
\begin{equation}
    \label{k}
    J_{\mathbf{x}}=-\kappa\nabla_{\mathbf{x}}T \, 
\end{equation}
where $J_{\mathbf{x}}$ is the heat flux along $\mathbf{x}$ direction
and $\nabla_{\mathbf{x}}\equiv\partial/\partial x$. The heat flux is
calculated by the energy per time per cross-sectional area that the
thermostats provide to the system. The temperature gradient is
calculated by dividing the TGNR in several slabs of length about 10
\AA, and determining the local temperature of each slab through the
average kinetic energy of the moving atoms over 10000 timesteps every
10000 timesteps. Figure \ref{fig02} shows a typical temperature
profile along the TGNR after the system reaches the steady state.

\section{Results}

\begin{figure*}[htb]
\centering
\includegraphics[width=0.9\linewidth]{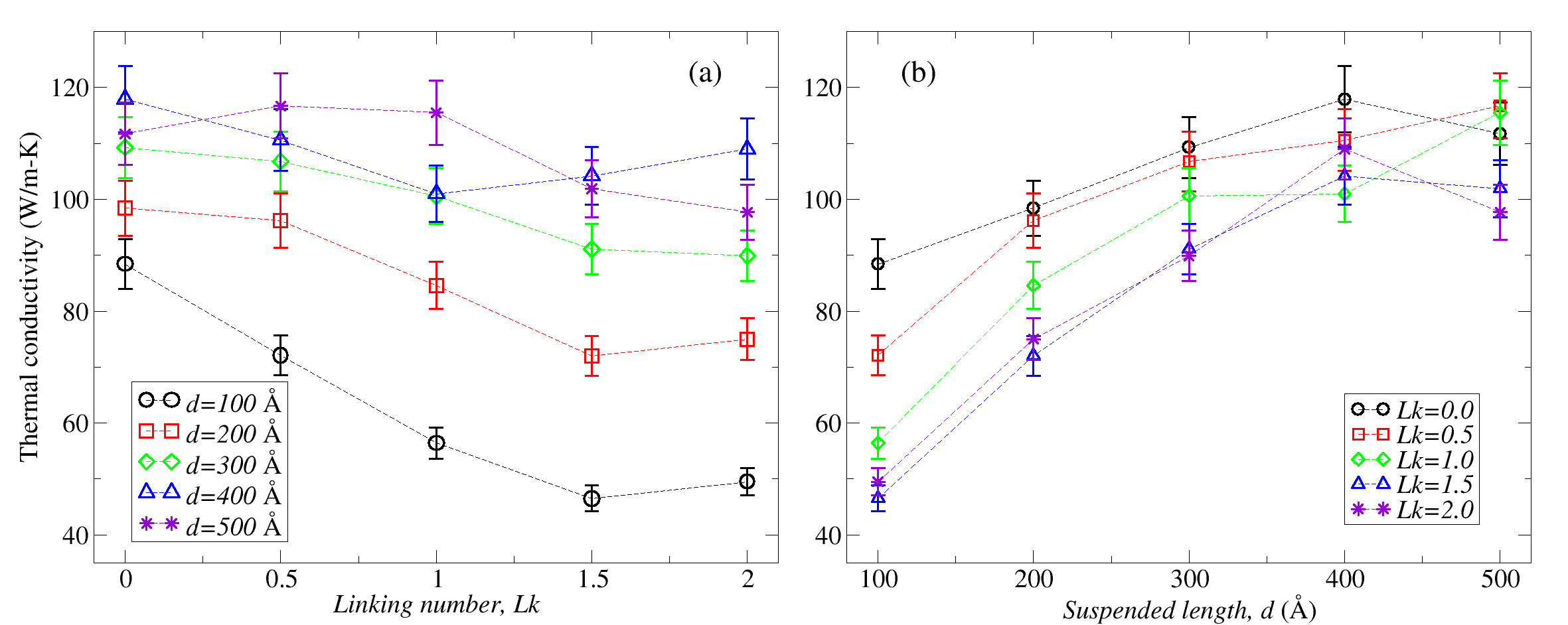}
\caption{(a) Thermal conductivity (TC) as a function of linking
  number, $Lk$, in each ribbon with increasing suspended length. (b)
  TC as a function of suspended length for ribbons with increasing
  $Lk$.  Dashed lines are just a guide to the eyes.}
\label{fig03}
\end{figure*}

Figure \ref{fig03}(a) presents the results for the TC as function of
$Lk$ for the structures at 300 K.  Each curve shows the TC for a given
value of suspended length, $d$, of the TGNR.  The curves have in
common the decrease of the TC with $Lk$ up to 1.5, after which the TC
remains approximately constant within the error bars, which represent
a 5\% uncertainty in all calculated conductivities.  The curves also
show that although the TC increases with increasing $d$, it seems to
converge, since the curves for $d\ge300$ \AA\mbox{} become closer to
each other than those for $d\leq300$ \AA\mbox{}.  As the linking
number, $Lk$, is determined by the number of turns initially applied
to the straight GNR, one might think that Figure \ref{fig03}(a)
represents the dependence of the TC of a TGNR on twist.  However, the
ability to change the suspended length, $d$, of the GNR, without
changing $Lk$, poses an extra complexity to the issue of dependence of
TC on twist.

Another form to see the complexity of the dependence of TC on twist
comes from the plot of TC as a function of the suspended length, $d$,
as shown in Figure \ref{fig03}(b).  The curves show that even for the
same value of $Lk$, the TC of a TGNR can change significantly.  This
observation confirms that, alone, $Lk$ cannot characterize the
dependence of TC on twist.  Figure \ref{fig03}(b) also allows us to
infer that the average rate of change of TC with $d$ roughly increases
with $Lk$, at least for $d\leq400$ \AA.

The above results indicate that the TC of a TGNR is not a simple
function of only one variable, the number of turns initially applied
to the GNR or $Lk$. The TGNR TC seems to require, at least, a second
parameter to appropriately describe its dependence on the geometric
features of the TGNR.  In order to find out which set of quantities
best suits this requirement, we propose, here, three possible
alternatives: (A) considering the TGNR suspended length, $d$, as
second parameter; (B) considering the pair $(Tw,Wr)$ parameters (or
equivalently, $Lk$ and one of them); and (C) considering the initial
twist density, $Lk/d$, as the second parameter.

\subsection{TC dependence on $Lk$ and $d$}
\label{secA}

\begin{figure}[htb]
\centering
\includegraphics[width=0.9\linewidth]{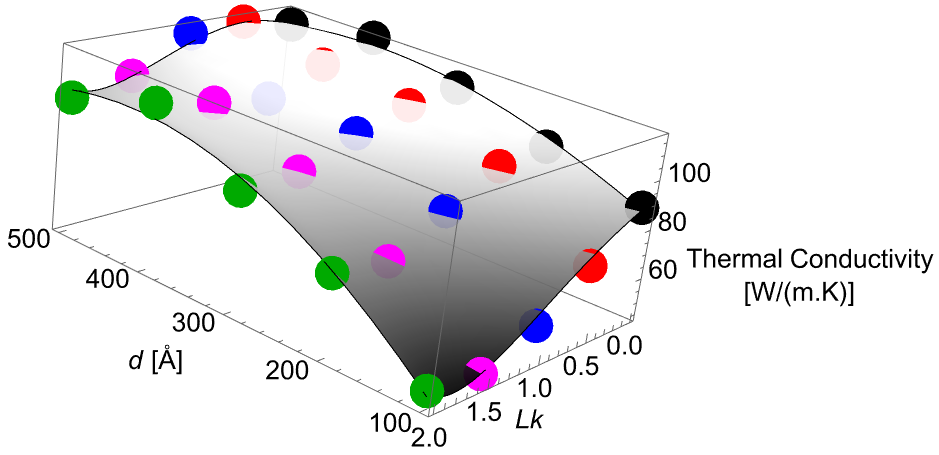}
\caption{Thermal conductivity (TC) as a function of both suspended
  length, $d$ and the Linking number $Lk$. Black, red, blue, magenta
  and green dots corresponds to the TC values of simulated TGNRs
  having $Lk=0$, 0.5, 1.0, 1.5 and 2.0, respectively. The gray surface
  is a fitting function of the TC points given by equation
  (\ref{eqfit}). See the text for details.}
\label{fig04}
\end{figure}

Figure \ref{fig04} illustrates the 3D distribution of values of the TC
of the TGNRs as functions of $d$ and $Lk$. The gray surface is the
result of an arbitrarily chosen fitting function TC = TC$(d,Lk)$,
given by:
\begin{dmath}
\label{eqfit}
\mbox{TC}(d,Lk) = 70.7806 - 28.7985\,d + 0.0439798\,d\ln{d}- 0.000383411\,d^2 - 0.098039\,d\,Lk - 8.59172\,Lk^2 - 0.000051652\,d^2Lk^2 + 1.08634\,Lk^5 \,,
\end{dmath}
\noindent where the parameters were determined by a nonlinear fitting. 

The functional form the the fitting function by itself is not so
important at the moment.  Different $d^n Lk^m$ terms could have been
added to the fitting equation with no significant difference in the
final result.  The point is that it is possible to obtain an empirical
analytical function for TC = TC$(d,Lk)$ from computational and/or
experimental data and, then, use it for future predictive purposes.
Here, Figure \ref{fig04} serves to reinforce the conclusion that the
TC of TGNRs cannot be described in a simplistic manner, solely in
terms of the number of initially applied turns or $Lk$.

\subsection{TC dependence on $Tw$ and $Wr$}
\label{secB}

The TC of the TGNRs can be correlated to their geometric features
twist, $Tw$, and writhe, $Wr$, instead of $d$ and $Lk$, because the
present simulations were conducted in such a way that the linking
number (or the initial number of turns applied to the straight GNR)
remained fixed.  Then, the C{\u a}lug{\u a}reanu-White-Fuller theorem,
eq. (\ref{lk}), can be used to distinguish groups of TC surfaces in a
$Tw\times Wr$ space, each one corresponding to a value of $Lk$.

\begin{figure}[htb]
\centering
\includegraphics[width=0.9\linewidth]{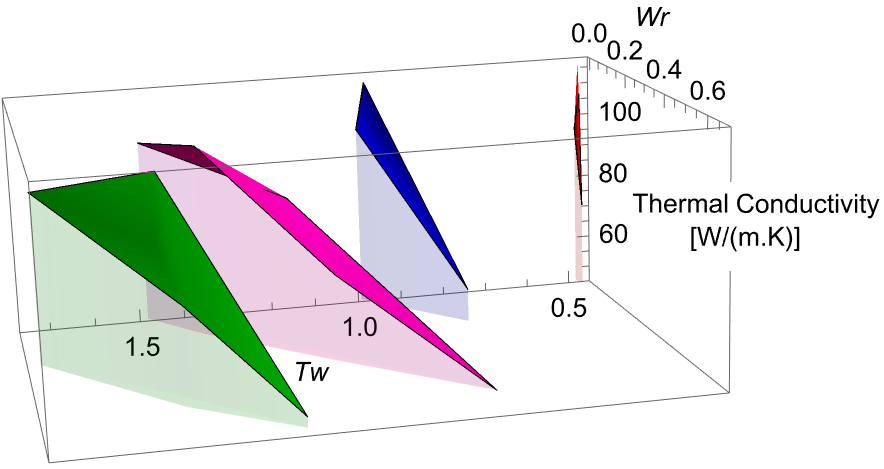}
\caption{Thermal conductivity as a function of both the twist, $Tw$
  and the writhe, $Wr$. Red, blue, magenta and green corresponds to
  TGNRs having $Lk=0.5$, 1.0, 1.5 and 2.0, respectively}
\label{fig05}
\end{figure}

Figure \ref{fig05} displays four non-zero constant-$Lk$ TC surfaces as
function of both $Tw$ and $Wr$ corresponding to the values of $d$ and
$Lk$ considered in the present study.  The area of the surfaces
increases with $Lk$ which reflects the ability of the TGNRs to convert
twist to writhe to release at least part of the torsional stress.  As
the surfaces do not intersect one another, each pair of geometric
coordinates, $(Tw,Wr)$ univocally characterizes the TC of a TGNR.

\begin{figure*}[htb!]
\centering \includegraphics[width=0.9\linewidth]{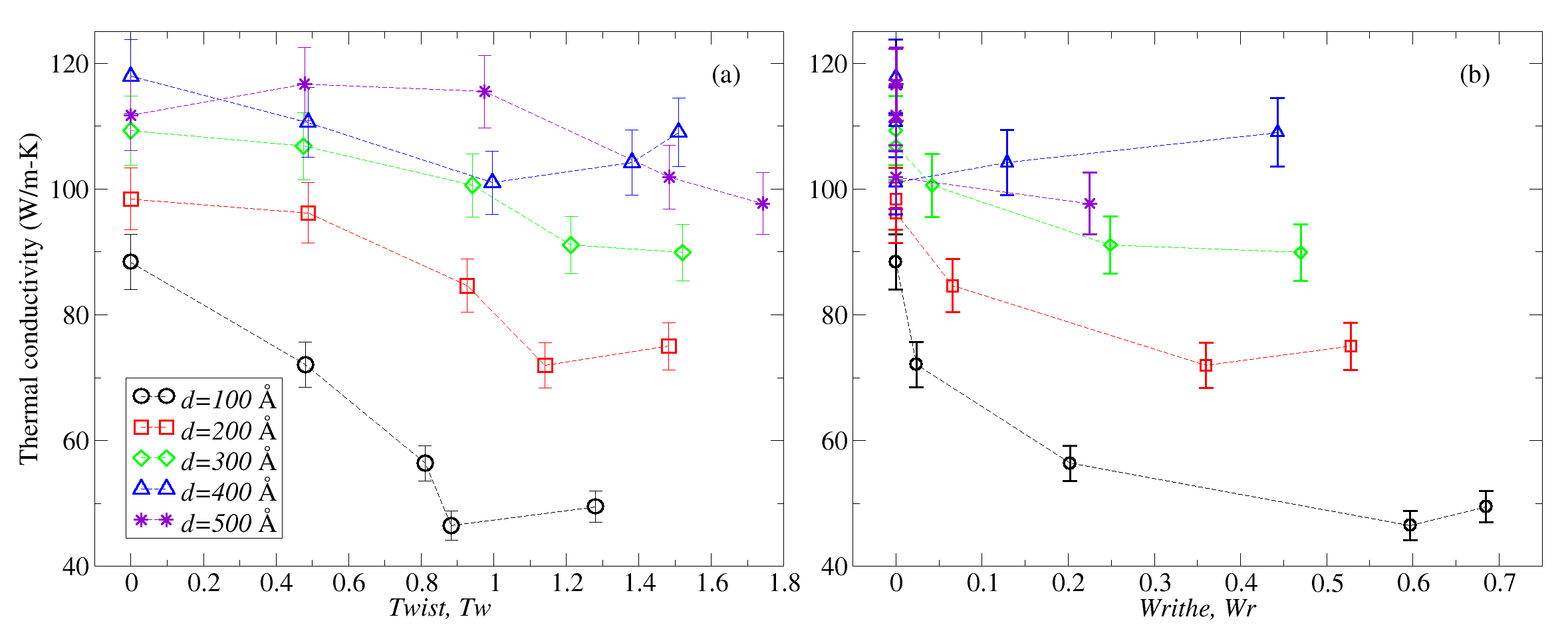}
\caption{(a) Thermal conductivity (TC) as a function of the twist
  $Tw$. (b) TC as a function of the writhe $Wr$. Dashed lines are just
  a guide to the eyes.}
\label{fig06}
\end{figure*}

Separated plots of TC versus $Tw$ and versus $Wr$, for different
values of suspended length, $d$, are shown in Figure \ref{fig06}.
They allow for a better observation of the complexity of the
dependence of TC of the TGNRs on their geometric parameters than
Figure \ref{fig05}.

Figure \ref{fig06}(a) shows that TGNRs with smaller suspended lengths,
$d$, present larger variations of the TC with the real twist, $Tw$.
In other words, as the value of $d$ increases, the TC becomes less
dependent on $Tw$.  This shows that the TGNR TC could not be described
uniquely even by the real twist, $Tw$.  In fact, the above result is
not unexpected.  As can be seen in the examples of TGNRs shown in
Figure \ref{fig01}, the suspended part of the structure becomes less
curved as $d$ increases.  This is a consequence of the decrease of the
linear twist density with increasing $d$ of the TGNR having the same
$Lk$.  Literature has shown that rods and ribbons become unstable when
the applied twist is such that the twist density becomes larger than a
critical value~\cite{sun98,goriely97,fonseca2006}.  Because of this,
we decided to also investigate the dependence of TC on the twist
density, as we will discuss in subsection \ref{secC}.

The curves in Figure \ref{fig06}(b) look different from those of panel
(a) but they are consistent and reflect the fact that $Tw$ and $Wr$
are, in fact, connected by Eq. (\ref{lk}).  In fact, Figure
\ref{fig06}(b) shows that for TGNRs with larger suspended length, $d$,
there is a larger number of TC points corresponding to $Wr<0.1$.  This
particular observation is coherent with the fact that increasing $d$
decreases the twist density of the TGNRs.  If the twist density is
small, the twisted but straight ribbon is a stable spatial
conformation.  In other words, the structure becomes less curved when
the twist density is low.  The points corresponding to values of
$Wr>0.2$ are those obtained for the largest values of $Lk$ considered
in the present study, or $Lk\geq1.5$.  Finally, Figure \ref{fig06}(b)
also shows that the TGNR TC cannot be described by only its writhe,
$Wr$.

\begin{figure}[htb]
\centering
\includegraphics[width=0.95\linewidth]{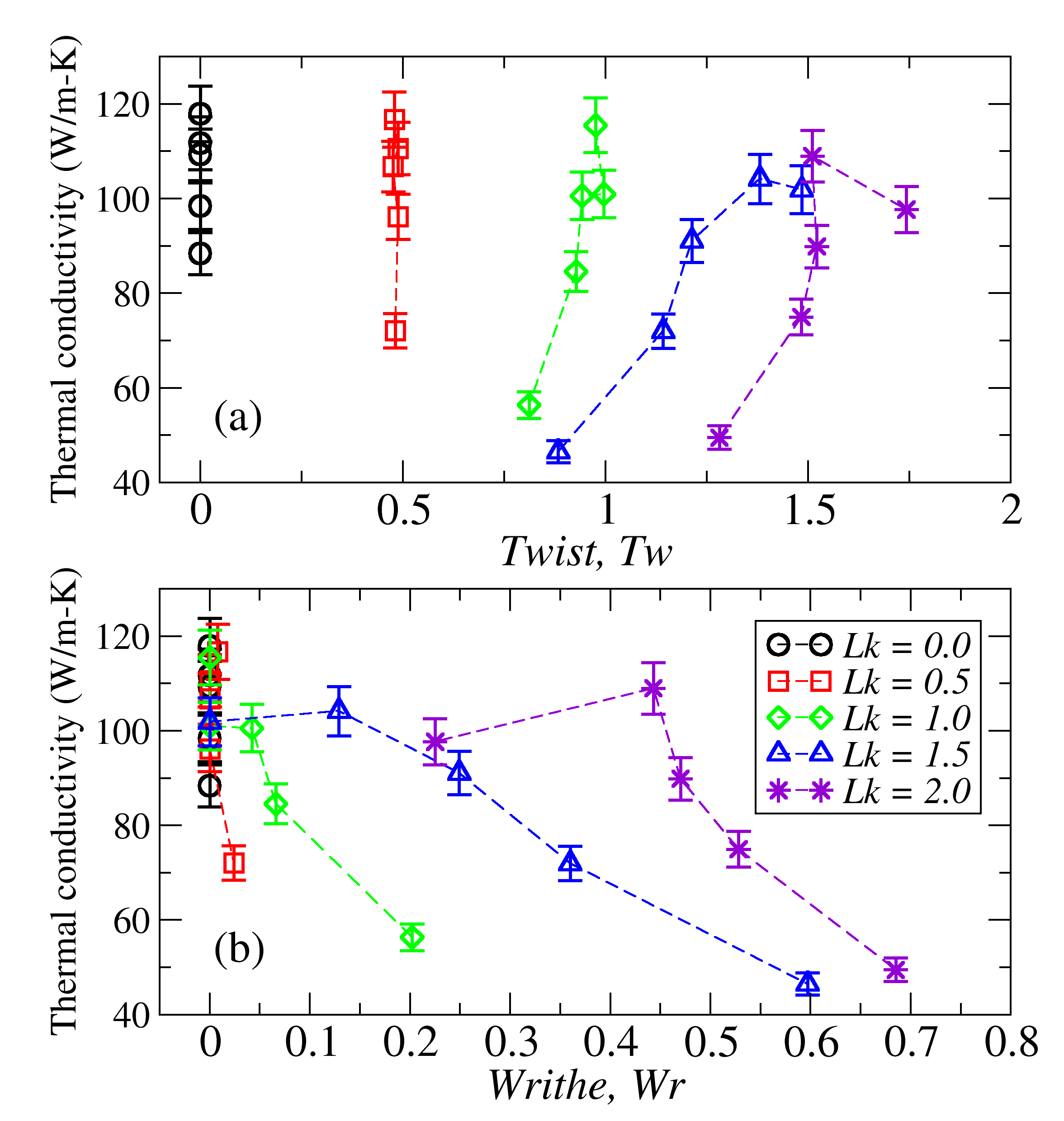}
\caption{The dependence of the thermal conductivity on the twist,
  $Tw$, (top panel) and writhe, $Wr$, (bottom panel) for each value of
  $Lk$. Dashed lines are just a guide to the eyes.}
\label{fig07}
\end{figure}

Figures \ref{fig06}(a) and (b) can be re-arranged if we group data
points by their $Lk$ values rather than $d$.  Indeed, Figure
\ref{fig07} shows TC as a function of $Tw$ (top panel) and $Wr$
(bottom panel), for each value of $Lk$.  Now, the curves display
additional results about the complex dependence of the TC of TGNRs on
their geometric features.  It can be seen that the rate of change of
TC with either $Tw$ or $Wr$, depends on $Lk$.  However, as $Lk=Tw+Wr$
is constant, $dTw=-dWr$ for the curves corresponding to the same
values of $Lk$.  Thus, for a given value of $Lk$, the rate of change
of TC with $Tw$ is equal in modulus to the rate of change of TC with
$Wr$.  For $Lk=0$, TC does not depend on $Tw$ or $Wr$, since there is
no twist and the different values of TC correspond only to different
values of $d$.  It is equivalent to say that $\Delta\mbox{TC}/\Delta
Tw\rightarrow\infty$ if $Lk,Tw,Wr\rightarrow0$.

However, for $Lk\neq0$, we observe that TC varies with either $Tw$ or
$Wr$. The difference amongst the curves is the rate of change of TC
with $Tw$ or $Wr$, $\Delta\mbox{TC}/\Delta Tw$, whose average values
are given in Table \ref{tab1}.  Although we have only a few values of
$\Delta\mbox{TC}/\Delta Tw$, and their values might have large
uncertainties, we can infer that they roughly decrease from a large
amount and converge with increasing $Lk$.

\begin{table}
  \caption{Average rate of change of TC with $Tw$,
    $\Delta\mbox{TC}/\Delta Tw$, for the curves shown in Figure
    \ref{fig07} and $Lk\neq0$.}
  \label{tab1}
  \begin{tabular}{cc}
    \hline
    \hline
    \hspace{1cm}$Lk$\hspace{1cm} &  $\Delta\mbox{TC}/\Delta Tw$ [W/(m-K)]  \\
    \hline
    0.5 & $11.15\times 10^{6}$ \\
    1.0 & 360.4 \\
    1.5 & 92.2 \\
    2.0 & 104.6 \\
    \hline
    \hline
  \end{tabular}
\end{table}

The above analysis confirms that the TC of TGNRs in realistic
conditions (large size GNRs, suspended and at different temperatures)
is much more complex than the simple consideration of its dependence
on the number of initially applied turns can deal with.  The
literature has only presented predictions for the dependence of TC on
twist, for zero K, non-writhed, small width TGNRs.

The study of the dependence of TC on both $Tw$ and $Wr$ is not
practical from the experimental point of view.  It is easier to record
the number of times the straight GNR was initially rotated and measure
the suspended length of the produced TGNR than defining two space
curves along the GNR and develop computational tools to extract its
points to compute the corresponding $Tw$ and $Wr$.  Therefore,
although the above results show the TC can be well characterized by
the pair $(Tw,Wr)$, they are not unique as shown in subsections
\ref{secA} and \ref{secC}.

\subsection{TC dependence on $Lk$ and the twist density}
\label{secC}

The analyses in the previous sections suggest one more attempt to
characterize the TC of TGNRs on just one quantity: the twist density.
In fact, there are two possible twist densities to consider, $Lk/d$
and $Tw/d$.  We will stick to the first one since, as mentioned in
Section \ref{secB}, it is easier for an experiment to measure $Lk$
then to measure $Tw$ in a TGNR.

\begin{figure}[htb]
\centering
\includegraphics[width=0.95\linewidth]{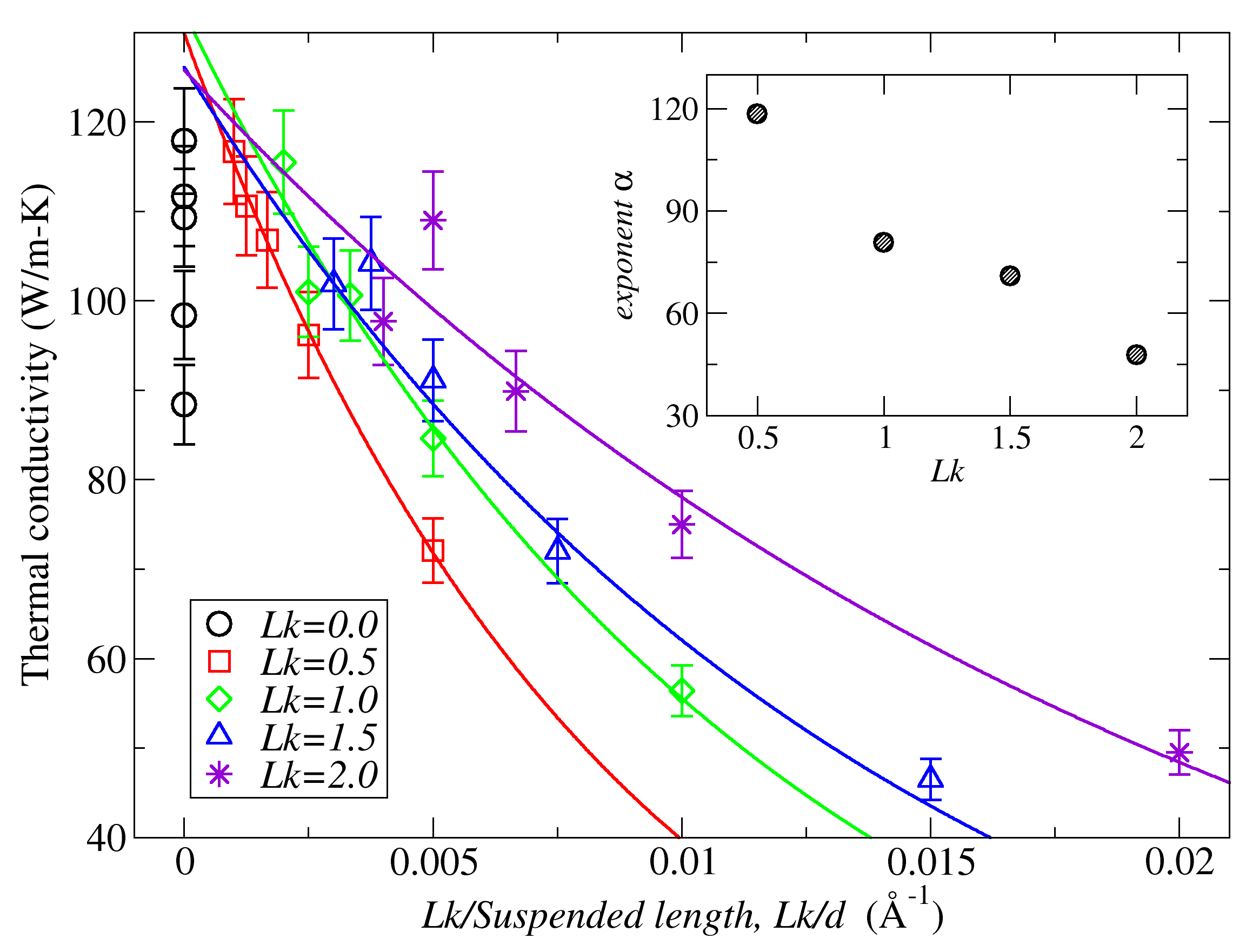}
\caption{Thermal conductivity as a function of the twist density,
  $Lk/d$. Points are results from MD simulations, and the curve is a
  fitting function given by equation (\ref{eqfit2}). Inset: fitting
  exponent $\alpha$ as a function of $Lk$.}
\label{fig08}
\end{figure}

Figure \ref{fig08} shows the TC of TGNRs as a function of $Lk/d$ for
different values of $Lk$.  It can be seen that each set of data points
corresponding to one value of $Lk$ seems to follow one particular
decaying curve.  We, then, fitted each one to the following equation:
\begin{equation}
\label{eqfit2}
f(x) = C e^{\left[-\alpha \left(x - x_0\right)\right]}, 
\end{equation}
where $C$ and $\alpha$ are constants that depend on $Lk$, and
$x_0=Lk/d_0$ is the smallest twist density of the set of data points
corresponding to the same $Lk$, with $d_0$ being the largest suspended
size of the TGNRs that, in our case, is 500 \AA.
The TC of TGNRs is, then, described by the following equation:
\begin{equation}
 \label{TCsecC}
\mbox{TC}(Lk,Lk/d)=C(Lk) e^{\left[-\alpha(Lk) \left( \frac{Lk}{d} -
    \frac{Lk}{d_0}\right)\right]} \, .  \end{equation}

\begin{table}
  \caption{Values of $C$ and $\alpha$ obtained from the fitting of the
    data points sets shown in Figure \ref{fig08}.}
  \label{tab2}
  \begin{tabular}{ccc}
    \hline
    \hline
    \hspace{1cm}$Lk$\hspace{1cm} & \hspace{1cm}$C$\hspace{1cm} & 
    \hspace{1cm}$\alpha$\hspace{1cm} \\
    \hline
    0.5 & 115.4 & 118.5 \\
    1.0 & 111.2 & 86.74 \\
    1.5 & 102.0 & 70.92 \\
    2.0 & 103.9 & 47.75 \\
    \hline
    \hline
  \end{tabular}
\end{table}

Table \ref{tab2} shows the values of $C$ and $\alpha$ obtained from
the fitting of the data points in the main panel of Figure
\ref{fig08}, for each value of $Lk$.  While $C$ is shown to be weakly
dependent on $Lk$, $\alpha$ seems to be a significant function of
$Lk$.
In fact, the inset of Figure \ref{fig08} shows that $\alpha(Lk)$
displays an approximately linear decreasing behavior with $Lk$.  The
meaning of this result is quite interesting.  As $C(Lk)\sim const.$,
one can conclude that larger values of $Lk$ lead to a weaker
dependence of the TC on the twist density $Lk/d$.  It is a remarkable,
and apparently contradictory, result because larger values of $Lk$
imply larger values of the twist density and, therefore, a stronger
dependence of TC on twist density. Thus, one would expect that
$\alpha(Lk)$ should increase and not decrease with $Lk$.  But the
richness of the twist-to-writhe phenomenon can help understand this
feature. As shown by Fonseca~\cite{fonsecaPRB2021}, a TGNR on top of
two separated substrates can present a twist-to-writhe transition,
where part of its twist is converted to writhe through curving and
curling in the space.  As this phenomenon decreases the torsional
stress on the nanoribbon, it decreases the torsional stress/strain
contribution to its TC.

While the dependence of the TC of TGNRs on $(Lk,d)$ is relatively
simple, equations similar to (\ref{eqfit}) are difficult to interpret
in physical terms.  However, although the dependence on $(Lk,LK/d)$ is
a bit more complicated than that on $(Lk,d)$, equation (\ref{TCsecC})
carries a simple and physically meaningful form of describing the TC
of TGNRs.

\section{Conclusion}

We have carried out fully atomistic molecular dynamics simulations of
TGNRs at 300 K, and obtained their thermal conductivity dependence on
geometrical parameters as twist, $Tw$, writhe, $Wr$, linking number,
$Lk$, size of the TGNRs suspended parts, $d$, and the twist density,
$Lk/d$.  The results showed that alone, the number of initially
applied turns to a straight GNR, $Lk$, cannot be considered the only
parameter that determines the TC of TGNRs.  We showed that the TC of
TGNRs can be a function of, at least, two parameters, and analysed
three sets of parameter pairs: $(Lk,d), (Tw,Wr)$ and $(Lk,Lk/d)$.
Even though each set of parameters can describe the TC of a TGNR, we
also showed that a simple and physically meaningful description can be
achieved with equation (\ref{TCsecC}), which describes the dependence
of TC on a twist density $(Lk,Lk/d)$.
In the present work, we were mostly concerned on showing that the
dependence of the TC on twist is not as simple as previous works have
suggested, which is probably related to the lack of experimental
studies on TGNRs.  We hope that the present analysis and findings will
stimulate further experimental investigations of TGNRs, including
their thermal transport properties.

\section*{Acknowledgements}

This work was financed by the Coordenação de Aperfeiçoamento de
Pessoal de N\'{i}vel Superior (CAPES) - Finance Code 001, Conselho
Nacional de Desenvolvimento Cient\'{i}fico e Tecnol\'{o}gico (CNPq),
S\~ao Paulo Research Foundation (FAPESP), Funda\c{c}\~ao de Amparo
\`{a} Ci\^{e}ncia e Tecnologia do Estado de Pernambuco (FACEPE), and
Financiadora de Estudos e Projetos (FINEP).  AFF acknowledges Grant
303284/2021-8 from CNPq, and Grants 2020/02044-9 and 2023/02651-0 from
FAPESP.  LFCP acknowledges Grant 0041/2022 from CAPES, 436859/2018,
313462/2020, 200296/2023-0 and 371610/2023-0 INCT Materials
Informatics from CNPq, APQ-1117-1.05/22 from FACEPE, 0165/21 from
FINEP, and the visiting professors program at Sapienza.  This work
used computational resources provided by “Centro Nacional de
Processamento de Alto Desempenho em São Paulo (CENAPAD-SP)” (project
proj937), and by the John David Rogers Computing Center (CCJDR) in the
Gleb Wataghin Institute of Physics, University of Campinas.


\end{document}